\documentclass[epj]{svjour}
\usepackage{graphics}
\usepackage{booktabs}
\usepackage{tabularx}
\usepackage[utf8]{inputenc}
\usepackage{lineno,hyperref}
\usepackage{subfigure}
\usepackage[disable]{todonotes}
\usepackage{tikz}
\begin{document}
\title{Toward inertial sensing with a monochromatic  $ 2^3 S $ positronium beam}
\author{Sebastiano Mariazzi \inst{1, 2}, Ruggero Caravita \inst{3}, Michael Doser  \inst{3}, Giancarlo Nebbia  \inst{4} \and Roberto S. Brusa \inst{1, 2}
}                     
\institute{Department of Physics, University of Trento, via Sommarive 14, 38123~Povo, Trento, Italy \and TIFPA/INFN Trento, via Sommarive 14, 38123~Povo, Trento, Italy \and Physics Department, CERN, 1211~Geneva~23, Switzerland \and INFN Padova, via Marzolo 8, 35131 Padova, Italy}
\date{Received: date / Revised version: date}
%
\abstract{
In this work, we discuss the possibility of  inertial sensing with positronium in the $2^3 S$ metastable state for the measurement of optical dipole, relativistic and gravitational forces on a purely leptonic matter-antimatter system. Starting from the characteristics of an available $2^3 S$ beam, we estimate the time necessary to measure accelerations ranging from $\sim10^5$ $m/s^2$ to 9.1 $m/s^2$ with two different inertial sensitive devices: a classical moiré deflectometer and a Mach-Zehnder interferometer. The sensitivity of the Mach-Zehnder interferometer has been estimated to be several tens of times better than that of the moiré deflectometer, for the same measurement time.\\
Different strategies to strengthen the $2^3 S$ beam flux and to improve the sensitivity of the devices are proposed and analyzed. Among them, the most promising are reducing the divergence of the positronium beam through 2D laser Doppler cooling and  coherent positronium Raman excitation from the ground state to the $2^3 S$ level. If implemented, these improvements promise to result in the time required to measure an acceleration of 9.1 $m/s^2$ of few weeks and 100 $m/s^2$ of a few hours.
Different detection schemes for resolving the fringe pattern shift generated on $2^3 S$ positronium crossing the deflectometer/interferometer are also discussed.
} 
\maketitle
\section{Introduction}
\label{intro}
 Positronium (Ps), the bound state of an electron and a positron with very short lifetime (125 ps in the singlet ground state, 142 ns in the triplet ground state), is one among the few neutral matter-antimatter/ pure antimatter systems experimentally accessible for testing symmetries and fundamental interactions on antimatter. Since its discovery in 1951 \cite{PhysRev.82.455}, Ps was used as a probe for nano/ mesoporosity in materials \cite{Gidley,Liszkay_2012,PhysRevLett.112.045501} and became a prime testing ground of bound-state Quantum Electrodynamics \cite{fee1993measurement,KARSHENBOIM20051,cassidy_review:18}. Ionic \cite{PhysRevLett.46.717,NAGASHIMA201495,michishio2016observation} and molecular \cite{Ps} Ps have been also observed and are currently being investigated.\\
At present, Ps plays a central role in one of the antimatter community's biggest efforts: measuring the gravitational force acting on antimatter bodies in the Earth’s gravitational field. The universality of the free-fall, i.e. the equivalence between inertial and gravitational masses, is a cornerstone of Einstein’s Equivalence Principle and General Relativity. Any observed discrepancy from the postulated equality would be an indication of new physics. On the one hand, Ps is being used by experiments in CERN’s Antiproton Decelerator as an (experimentally demonstrated \cite{PhysRevLett.93.263401}) intermediate tool to produce antihydrogen via a charge-exchange reaction with antiprotons. The goal of these experiments is to probe gravity using antihydrogen \cite{aegis_grav:12,Perez_2012}. Both a classical deflectometer \cite{moire} and an interferometer \cite{ALPHA} have been proposed for inertial sensing measurements with antihydrogen. On the other hand, several experimental schemes to directly measure gravity with long-lived Ps beams have been proposed (thus probing the gravitational interaction with a purely leptonic matter-antimatter system), either by letting Ps atoms free-fall in a drift tube and observing their vertical displacement with position-sensitive detector \cite{mills_leventhal:02} or by adopting a matter-wave atom interferometer \cite{oberthaler_ps:02}.\\ Ps has been also proposed as a probe to investigate forces other than gravity \cite{oberthaler_ps:02}. Indeed thanks to its low mass and high magnetic moment \cite{cassidy_review:18}, Ps is expected to be the ideal system to study relativistic forces like the Anandan force \cite{ANANDAN1989347}. Moreover, one can exert a force on Ps via electric dipole interactions with far-detuned laser light (optical dipole force). Optical dipole forces can exceed by up to several orders the magnitude of the gravitational force \cite{dipole_force}. 

\begin{table*}
\centering
\caption{Efficiency of $e^+ \rightarrow 2^3 S$ positronium conversion according to presently available techniques.  Bunches of $10^7 e^+$ every $\sim$40s have been considered for the final estimation of $2^3 S$ production rate.}
\label{tab:1}       
\begin{tabular}{lccc}
\toprule
 Process & Efficiency & $2^3 S$ production \\
\midrule
$e^+$/Ps conversion efficiency & $\sim$0.3 \cite{mariazzi2010prb} &  \\
$1^3 S \rightarrow 3^3 P\rightarrow 2^3 S$ excitation efficiency & $\sim$0.015  \cite{PhysRevA.99.033405} & $10^7$$ *  0.3 * 0.045$ $\sim$ $1.3 * 10^5$ \\
$3^3 P\rightarrow 2^3 S$ stimulated decay  & x 3 \cite{arXiv:1904.09004} &  per bunch\\
overall $ 2^3 S$ production efficiency  & $\sim$0.045 (0.015 x3) &\\
\bottomrule
\end{tabular}
\end{table*}

 Such force-sensitive inertial experiments require a long-lived Ps beam. A widely adopted approach to increase the short lifetime of Ps is to excite it to Rydberg levels (n=15-30) with a two-step laser excitation scheme, i.e. either using n=2 \cite{cassidy2012efficient} or n=3 \cite{aegis_neq3:16} as intermediate levels. This approach is beneficial for the long lifetimes that can be achieved ($>$100 $\mu$s), yet exhibits some potential limitations in view of measuring tiny forces (such as gravity). Rydberg Ps has a non-negligible electrical polarizability (which scales as $n^6$), making it sensitive to stray electric field gradients \cite{cassidy_review:18,mills_leventhal:02}, and spontaneously decays towards lower excited states populating a large number of sublevels. Exciting Ps to circular Rydberg states would mitigate this issue \cite{Jones_2016}, at the price of flux and a much more complex laser excitation stage \cite{mills_leventhal:02}.  An alternative pathway consists in laser-exciting the atoms to the long-lived $ 2^3 S $ metastable level, either by a two-photon transition \cite{fee1993measurement} or using a pulsed mixing electric field \cite{PhysRevA.95.033408} or by spontaneous decay from n=3 \cite{aegis_meta:18}. In the absence of strong electric fields, Ps in this level cannot decay optically (due to dipole selection rules) and has a relatively long annihilation lifetime (1.14 $\mu$s).\\
In this paper, we investigate the possibility to perform force-sensitive inertial experiments with $ 2^3 S $ Ps. We will discuss the time needed (that is related to the characteristics of the $ 2^3 S $ Ps beam) to perform such experiments because it represents one of the main limitations with the present technologies. The work is organized as follows. In Sec.2, the characteristics of the monochromatic $ 2^3 S $ Ps source currently available (intensity, $ 2^3 S $ velocity and angular divergence) are summarized \cite{aegis_meta:18,PhysRevA.99.033405,arXiv:1904.09004}. In Sec.3, the performances of a moiré and a Mach-Zehnder interferometer are compared. The feasibility of force-sensitive inertial experiments on $ 2^3 S $ is discussed and the required time to measure forces of different intensity on this system is estimated. Sec.4 is devoted to the analysis of the detection schemes of the fringe pattern generated by the deflectometer/interferometer. Finally, in Sec.5, we discuss possible improvements to enhance the intensity and quality of the $ 2^3 S $ beam and the sensitivity of the interferometer.

\section{$2^3 S$ positronium beam: state of the art}
\label{sec:1}
\begin{figure}[htb]
\centering
\includegraphics[width=0.45\textwidth]{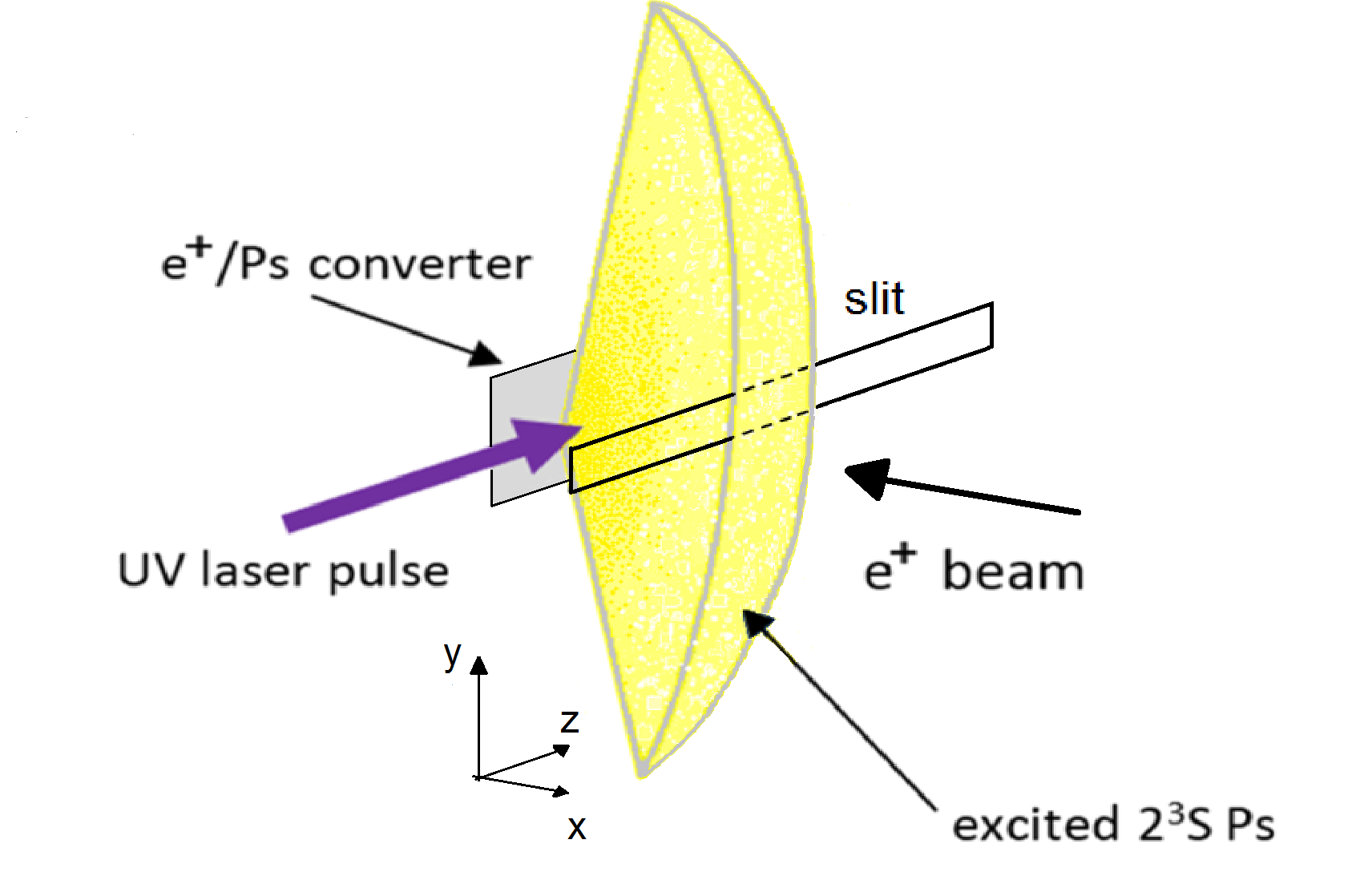}
\caption{
Schematic of the expansion in vacuum of the $ 2^3 S $ Ps produced according to Ref.\cite{PhysRevA.99.033405}. The partially-dashed rectangle indicates a slit that can be used to select a collimated fraction of the atoms in the vertical direction. The x-y-z axes are reported (see text).}
\label{Fig.1}
\end{figure}

Ps in the  $ 2^3 S $ state can be produced via spontaneous decay from the  $ 3^3 P $ state previously populated from the ground state  $ 1^3 S $ with a 205 nm UV laser pulse \cite{aegis_meta:18,PhysRevA.99.033405}.\\ The use of an efficient positron trapping and storage technology \cite{PhysRevA.46.5696,danielson2015} allows to produce bunches of around $ 10^7 $ positrons compressed in less than 10 ns \cite{cassidy2006accumulator, aegis_nimb:15}. If 50 mCi ${}^{22}$Na sources are employed, these bunches can be implanted every $ \sim $ 40 s in $e^+$/Ps nanochannelled silicon converters from which more than 30 $\%$ of implanted positrons reemerge as Ps (when $e^+$ are implanted with an energy of 3-5 keV) \cite{mariazzi2010prb, mariazzi2010prl}. Ps emission is expected to be roughly isotropic \cite{PhysRevA.99.033405}. A fraction of the emitted Ps can then be excited to the $ 3^3 P $ state \cite{aegis_neq3:16}. The subsequent $ 3^3 P \rightarrow 2^3 S $ spontaneous decay has been found to occur with a branching ratio of $\sim 10 \%$, in good agreement with theoretical expectations \cite{aegis_meta:18,PhysRevA.99.033405}. The short duration of the laser pulse used to excite Ps to the $ 3^3 P $ state allowed selecting a fraction of the Ps cloud with a quasi-monochromatic velocity distribution \cite{PhysRevA.99.033405}. Moreover, by changing the laser pulse delay with respect to the positron implantation instant, it has been possible to address Ps populations emitted with different velocities and thus a tuning of the $2^3 S$ velocity has been obtained \cite{PhysRevA.99.033405}. With a laser pulse delay of 20 ns, we observe a velocity distribution with a mean of $10^5$ m/s and a standard deviation of less than $10^4$ m/s. The $2^3 S$ production efficiency has been found to be up to 1.5 $\%$ of the overall number of Ps emitted by a nanochanneled silicon target \cite{aegis_meta:18,PhysRevA.99.033405}. A further factor $\sim$ 3 of increase in the $ 2^3 S $ production efficiency has been achieved by employing a 1312 nm laser pulse to stimulate the transition from the $ 3^3 P $ to the $ 2^3 S $ level \cite{arXiv:1904.09004}. As summarized in Tab.1, combining all these currently available techniques, around $1.3 * 10^5$ monochromatic  $ 2^3 S $ Ps atoms can be produced every $\sim$ 40s. The bandwidth of the 205 UV laser pulse (120 GHz), used to populate the $ 3^3 P $ level, is narrower than the Doppler profile of Ps emitted from the converter \cite{PhysRevA.99.033405}. As a consequence, the UV pulse performs a Doppler selection on the Ps along the laser propagation line (z axis in Fig. \ref{Fig.1}). 

If Ps atoms with average velocity of $10^5$ m/s are considered, the Doppler selected $ 2^3 S $ Ps expand within a spherical wedge of around $\pm 14^{\circ}$ with respect to the normal plane to the target (x-y plane in Fig. \ref{Fig.1}) \cite{PhysRevA.99.033405}. By using a slit, it is possible to select a collimated fraction  in a given direction (y direction in Fig.1) of this spherical wedge, at the price of a reduction in the flux. In Fig.2 the number of $ 2^3 S $ still alive at a given distance from the source in a beam selected with a slit fixing the vertical divergence to 17 mrad (5.5 mrad) is reported. The numbers are for a source of $ 1.3 * 10^5 $ $ 2^3 S $ Ps atoms (see Tab.1).

\begin{figure}[htb]
\centering
\includegraphics[width=0.9\linewidth]{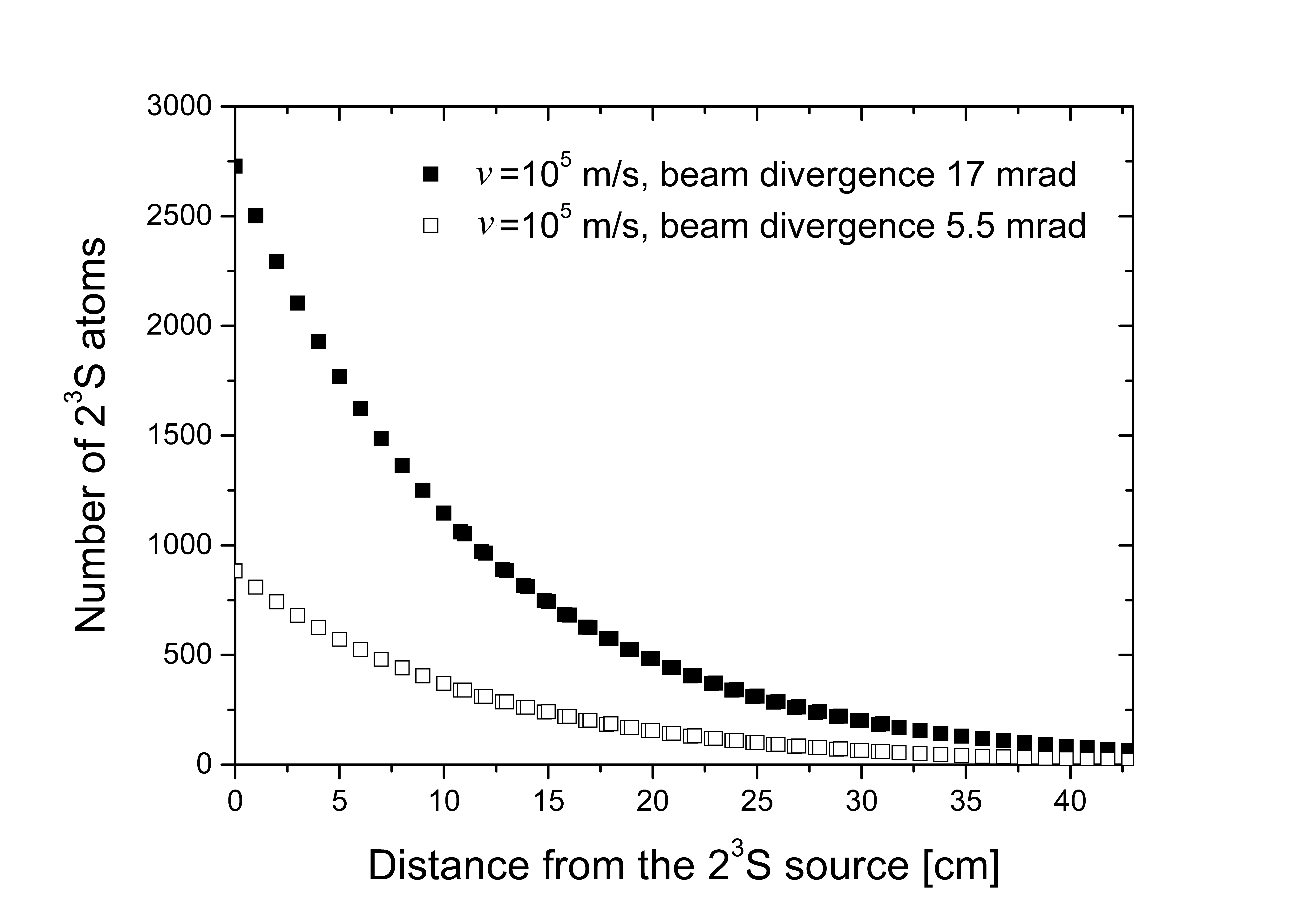}
\caption{
Number of surviving $ 2^3 S $ Ps atoms per bunch as a function of the travelled distance from the $e^+$/Ps converter in a beam with a vertical divergence of 17 mrad (black squares) and 5.5 mrad (open squares). A source intensity of $1.3 * 10^5$ $ 2^3 S $ atoms and a velocity of $10^5$ m/s for the $ 2^3 S $ atoms in a $\pm14^{\circ}$ wedge have been assumed.}
\label{Fig.2}
\end{figure}

\section{Deflectometry/interferometry systems}
\label{sec:1}

We discuss here deflectometry/interferometry inertial sensitivity devices composed of two gratings and a detector. As an example, a classical device is reported in Fig.3.

\begin{figure}[ht]
\centering
\includegraphics[width=\linewidth]{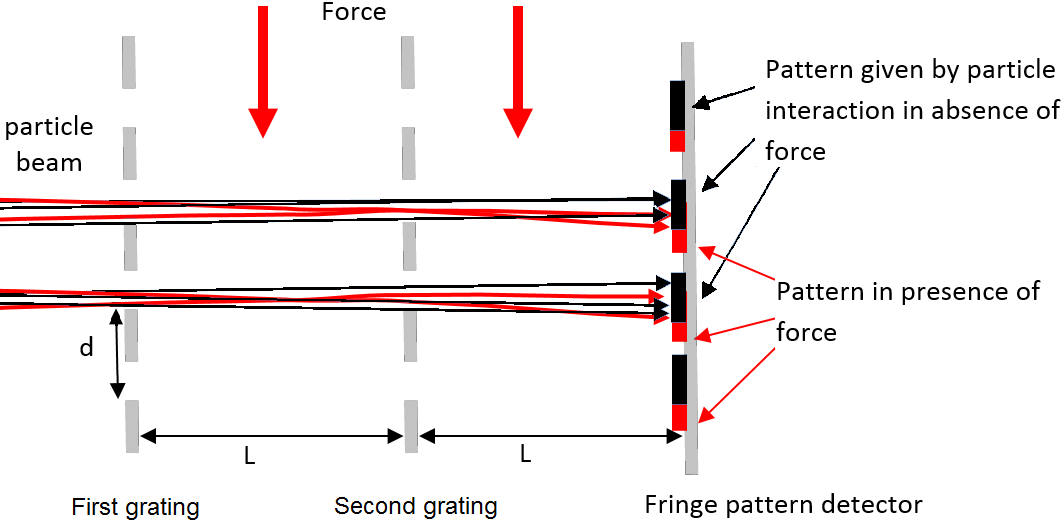}
\caption{
Schematic of a classical two-grating deflectometry system with periodicity $d$.}
\label{Fig.3}
\end{figure}

The gratings can be either physical or light gratings and they can stop and/or deflect some of the incident particles. As a consequence, the particles crossing both the grids generate a periodic pattern at the detector position. The presence of an eventual force acts on the position of this pattern. Thus, the presence and the intensity of a force can be determined by studying the position of the pattern.

\subsection{ Moire deflectometer}
\label{sec:2}

The moiré deflectometer is a classical device where the gratings are physical gratings with a periodicity $d$ much larger than the de Broglie wavelength ($\lambda_{dB}$) of the test particles \cite{PhysRevA.54.3165}.

\begin{equation}%
	\lambda_{dB} ~=~ \frac{h}{mv} \, ,
	\label{eqn:sdef}
\end{equation}

where $m$ is the inertial mass of the test particle, $v$ its velocity and $h$ the Planck constant. The classical trajectories defined by the two gratings lead to a fringe pattern with the grating periodicity at the longitudinal distance $L$ from the second grating where the detector is positioned \cite{moire}. If the transit time of the particles through the device is known, absolute force measurements are possible by employing Newton’s second law of dynamics. As indicated in Fig.3, the position of the moiré pattern is shifted in presence of a force with respect to the geometric shadow by:

\begin{equation}%
	\Delta_{y} ~=~ \frac{F}{m}t^2 ~=~ at^2\, ,
	\label{eqn:sdef}
\end{equation}

where $F$ represents the force component along the grating period, $a$ is the acceleration and $t$ is the time of flight between the two gratings spaced a length $L$ apart. The time $t$ is called interrogation time. Thus, in the case of a monochromatic beam with known velocity $v$, $t$ is simply $L/v$. 
The sensitivity of the system (i.e. the minimum detectable acceleration, $a_{min}$) is given by:

\begin{equation}%
	\ a_{min} ~=~ \frac{d}{2\pi V t^2 \sqrt{N}}\, ,
	\label{eqn:sdef}
\end{equation}

where $N$ is the number of detected particles and $V$ is the visibility of the fringes \cite{moire}. For particles with very long lifetimes, by increasing the distance $L$ between the gratings, it is possible to enhance the sensitivity simply by increasing the interrogation time $t$. However, for particles with a limited lifetime -like metastable Ps- the number of particles reaching the detector scales as Eq.(4):

\begin{equation}%
	\ N ~=~ N_{0}T e^{-\frac{2t}{\tau}}\, ,
	\label{eqn:sdef}
\end{equation}

where $ N_{0}$ is the number of particles at the entrance of the device, $\tau$ is the particle lifetime (1142 ns for $ 2^3 S $ Ps) and $T$ is a factor of transparency of the gratings. A value of 1 corresponds to the case that no particles are stopped. Combining Eq.(3) and Eq.(4), one can estimate the behavior of the minimum detectable acceleration for $ 2^3 S $ Ps as a function of the interrogation time (Fig.4).

\begin{figure}[htp]
\centering
\includegraphics[width=\linewidth]{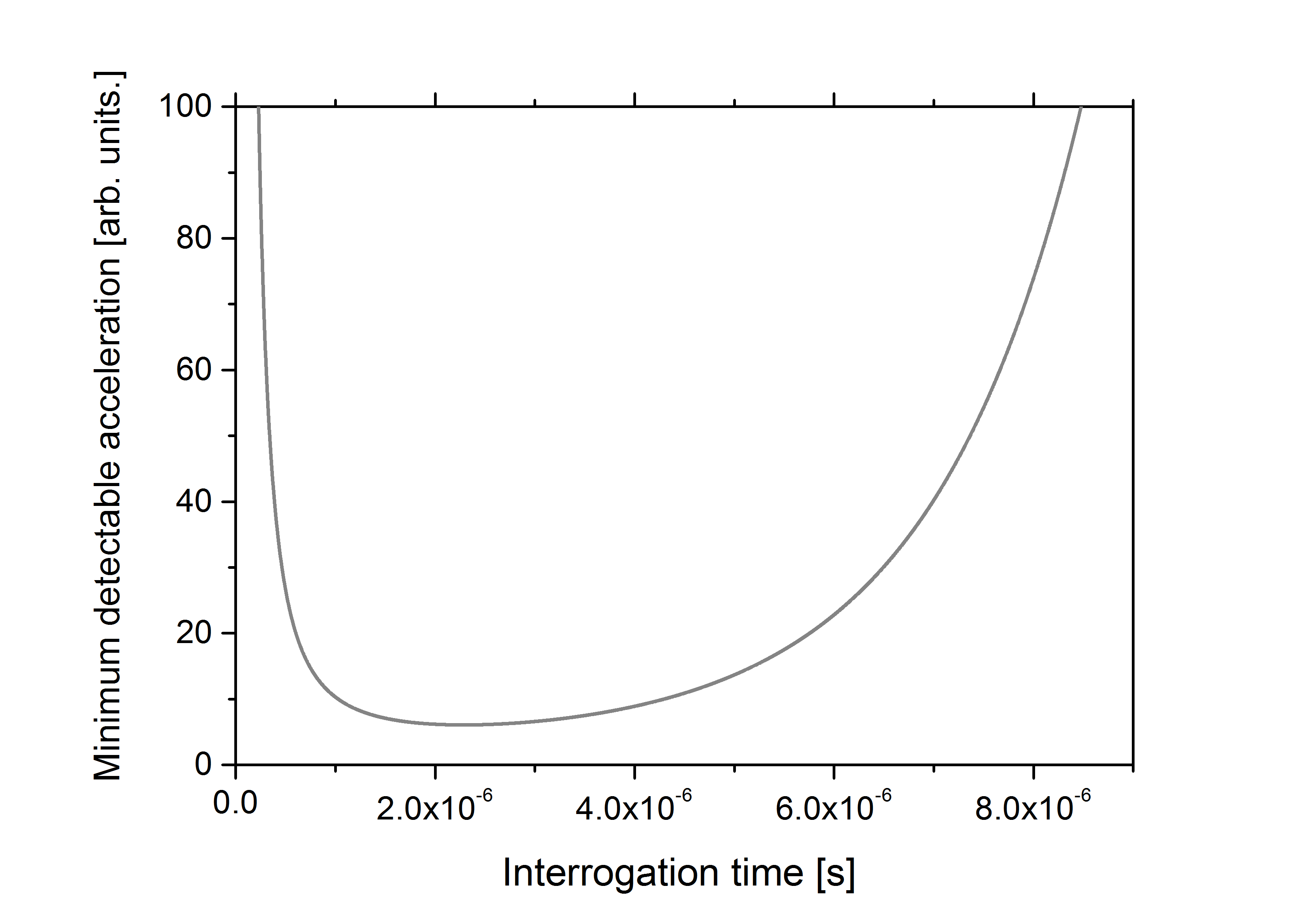}
\caption{
Sensitivity estimate for a  $ 2^3 S$ interferometer. The graph shows the behavior of the minimum detectable acceleration in arbitrary units as a function of the interrogation time.}
\label{Fig.4}
\end{figure}

The plot shows that the minimum acceleration sensitivity is reached for $\sim 2 \mu s$ i.e. 2 times the $ 2^3 S $ lifetime. This means that, once the velocity of the beam is defined, the best distance between the gratings is also set to $L \sim v * 2 \mu s$. 
The moiré deflectometer offers the advantage of working even for a divergent source of particles \cite{phd_braunig:14} allowing the use of a relatively large solid angle of the beam without sacrificing particles. On the other hand, the number of particles reaching the detector is limited by the transparency of the material gratings that is far from 1 \cite{phd_braunig:14}. Moreover, the dimension of the grating period sets an important constraint on the minimum measurable acceleration with this device [Eq.(3)]. With a divergence of 17 mrad (consistent both with the geometry of our experimental set-up of the $ 2^3 S $ beam \cite{aegis_nimb:15} and the angular acceptance of a moiré deflectometer \cite{moire}), with a grating period of the order of 40 $\mu$m (a slit width of 12 $\mu$m) and a transparency of the gratings of  $\sim$0.3, a visibility of the fringes of $\sim$0.3 can be obtained \cite{moire,phd_braunig:14}. These grating dimensions are currently feasible \cite{moire}.

\begin{table*}
\centering
\caption{Comparison of the minimum measurable acceleration in 4.5 hours, 2.6 weeks and 60 months with a moiré deflectomer and a Mach-Zehnder interferometer by employing the present $e^+$ storage technology, $e^+$/Ps converters and tested strategies for the production of monochromatic $2^3 S$.} 
\label{tab:2}
\begin{tabular}{cccccc}
\toprule
deflectometer/ &  angular  & Number of $2^3 S$  & $a_{min}$  &  $a_{min}$  & $a_{min}$  \\
interferometer&divergence&at the entrance of the device &in 4.5 hours&in 2.6 weeks &in 60 months\\
&& per shot&($\sim$4 $10^2$ shots)&($\sim$4 $10^4$ shots)&($\sim$4 $10^6$ shots)\\
\midrule
moiré                                      & 17 mrad                      & $\sim$2100	                                                                    & $\sim6.1$  $10^3$ g	            & $\sim 6.1$  $10^2$ g	                                     &                         $\sim$61 g \\
Mach-Zehnder                        &	5.5 mrad	          & $\sim$ 680	                                                                               & $\sim$ $10^2$ g	            & $\sim$ 10 g	                                                &                         $\sim$ 1 g\\
\bottomrule
\end{tabular}
\end{table*}

With the characteristics of the $ 2^3 S $ beam reported in section 2, taking atoms with a velocity of $10^5$ m/s (i.e. $L \sim v * 2 \mu s$=20 cm), the vertical divergence indicated above and the first grating at 3 cm from the target, a minimum acceleration of 6.1 $10^3$ g, $6.1 * 10^2$ g and 61 g can be measured  in $4 * 10^2$ shots ($\sim$4.5 hours), $4 * 10^4$ shots ($\sim$2.6 weeks of measurement) and $4 * 10^6$ shots (60 months) [Eq.(2) and Eq.(3)]. The sensitivity estimations are reported in Tab.2.\\
The discussion above does not take into account the presence of possible systematics such as external field gradients or longitudinal asymmetries and rotational misalignements of the gratings. The effect of such eventual systematics would be to degrade the quality of the pattern, reducing the visibility $V$ of the fringes \cite{PhysRevA.96.063604} and consequently increasing the measurement time [Eq.(3)]. As our experiment can be done in a free field region (see Ref.\cite{aegis_nimb:15,PhysRevA.99.033405} for details), the effect of field gradients is expected to be limited. On the opposite, the precision in the alignement of the gratings is of great importance. According to Ref.\cite{PhysRevA.96.063604}, the effect of a longitudinal displacement that asymmetrize the position of the two gratings and/or the fringe pattern plane as well as the effect of a rotational misalignement between the gratings are directly proportional to the periodicity $d$ and inversely proportional to the divergence of the beam. For the deflectometer and the beam considered in this section, a visibility of the fringes higher than $\sim$0.15 can be obtained with a longitudinal misplacement of the gratings smaller than $\sim$0.6 mm and a rotational misalignement smaller than $\sim$3 mrad \cite{PhysRevA.96.063604}.

\subsection{Mach-Zehnder interferometer}
\label{sec:2}

An alternative scheme is constituted by the Mach-Zehnder interferometer (see for example \cite{Baudon,Miffre,Cronin}). Such as scheme has been proposed for positronium in Ref.\cite{oberthaler_ps:02}. As depicted in Fig.5, it requires generating standing waves of light where the diffractive optical elements are thicker than the Talbot length ($L_{Talbot}$):

\begin{equation}%
	\ L_{Talbot} ~=~ \frac{2d^2}{\lambda_{dB}}\, ,
	\label{eqn:sdef}
\end{equation}

for the considered Ps with a velocity of $10^5$ m/s, $\lambda_{dB}=  h/(m v)$=3.6 nm.  In this configuration an atom incident at the Bragg angle $\theta_B$ traverses more than one grating plane and a fraction of the incident particles is deviated with an angle of:

\begin{equation}%
	\ \theta ~=~ 2\theta_B ~=~ 2\frac{\lambda_{dB}}{2d}\, ,
	\label{eqn:sdef}
\end{equation}

A standing light wave with wavelength $\lambda$ leads to a periodic potential for the atoms with $d=\lambda/2$. Following the original scheme proposed in Ref.\cite{oberthaler_ps:02}, where $\lambda$ =1312 nm is used, the diffraction angle $\theta$ = 5.5 mrad \cite{note}.  In order to be able to distinguish the interference fringes, the angular collimation of the beam has to be smaller than this diffraction angle given by the standing light wave of 5.5 mrad. 

\begin{figure}[ht]
\centering
\includegraphics[width=1 \linewidth]{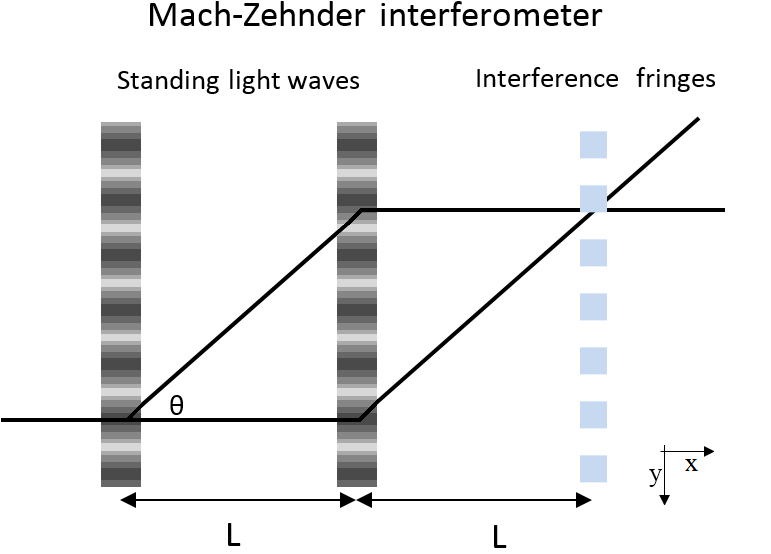}
\caption{
 Schematic of a matter wave Mach–Zehnder interferometer. Two standing light waves act like diffractive optical elements deflecting a fraction of the beam and creating the interference fringes.}
\label{Fig.5}
\end{figure}

The expected number of $ 2^3 S$ Ps in the slit with a vertical divergence of 5.5 mrad as a function of the distance from the target is shown in Fig.2. If the first light grating is placed at 3 cm from the source, around 680 $ 2^3 S$ Ps atoms are expected to reach the entrance of the interferometer with vertical divergence of 5.5 mrad. The reduction of the beam flux, required by the Mach-Zehnder interferometer, negatively affects the sensitivity of the device [Eq.(3) and Eq.(4)]. However, this effect is compensated by the reduction of the grating period $d$ and -more importantly- by the very high transparency of the light gratings that approaches unity. As a result, this device would allow to detect an acceleration of $10^2$ g, 10 g and 1 g in $4 * 10^2$  shots ($\sim$4.5 hours of measurement), in $4 * 10^4$  shots ($\sim$2.6 weeks of measurement) and $4 * 10^6$ shots ($\sim$ 60 months of measurement), assuming a visibility of the fringes of 0.3 and a dimension of the device like in section 3.1 (for the feasibility of a Mach-Zehnder interferometer with $L$ of the order of tens of centimeters see, for example, Ref. \cite{Giltner}). These estimates are reported in Tab.2. In comparison to the moiré deflectometer, the Mach-Zehnder interferometer achives several tens of times better sensitivity for the same measurement time.\\
The reduction of the grating period, in the interferometer with respect to the moiré deflectometer, will require a better precision in the alignement of the standing waves of light in order to preserve the visibility of the fringes. With the interferometer and the beam considered in this section, a visibility higher than $\sim$0.15 will require a longitudinal asymmetry of the gratings smaller than $\sim$30 $\mu$m and a rotational misalignement smaller than $\sim$150 $\mu$rad \cite{PhysRevA.96.063604}.

\section{Fringe pattern detection}
\label{sec:1}

Two different approaches can be used to detect the fringe pattern of the deflectometer/interferometer. The first method consists in resolving the pattern with a position sensitive detector. The second one is to probe the fringe pattern with a third grating followed by a detector with no spatial resolution, measuring the particle flux \cite{phd_braunig:14}.\\
The limits of the application of a position sensitive detector in the case of $2^3 S$ Ps are evident from the comparison of Eq.(2) and Fig.4. Indeed, the figure tells us that the minimum detectable acceleration can be obtained for interrogation times of the order of 2-5 $\mu$s. According to Eq.(2), this means that shifts of $\Delta y <$ 0.3 nm (alternatively, 3 $\mu$m) should be resolved to detect an acceleration of $\sim$g (alternatively, $\sim 10^4$ g). Thus far, the best resolution in the determination of the annihilation position of a Ps atom on a plane is of the order of 90  $\mu$m \cite{MCP}. This resolution has been achieved with a detector based on the ionization of Ps in a strong homogeneous magnetic field and the detection of the photo-positrons with a microchannel plate (MCP)+phosphor screen assembly \cite{MCP}. In the present context, a position sensitive detector could be realized by employing a continuous laser to photoionize $2^3 S$ Ps in the plane where the fringes have to be resolved. Extracted $e^+$ can then be guided by a homogeneous magnetic or electric field (taking care that the fringe field does not penetrate in the deflectometer/interferometer) towards the MCP assembly. Thanks to the guiding field, the charges conserve the information of the Ps impact position within the photoionizing plane \cite{MCP}. The limits to the spatial resolution are the cyclotron radius of the charges in the magnetic field, that can be reduced by increasing the intensity of the magnetic field (for a free $e^+$ with an energy of some tens of meV, the cyclotron radius is of the order of 1 $\mu$m in a field of 1 T), and -more importantly- the spatial resolution of the MCP assembly. A reduction of the spatial resolution to values lower than 10 $\mu$m could be obtained by employing MCPs with small pores and by replacing the phosphor screen with Timepix chips (see for example Ref.\cite{PACIFICO201612}), which allows to implement methods of centroid reconstruction \cite{Tremsin_2018}. Even better spatial resolutions, in such a Ps detection scheme, could be obtained by using an emulsion detector [see for example \cite{Aghion_2013}] placed downstream of the photoinizing plane. Photo-detached positrons can be accelerated to the emulsion and imaged with a spatial resolution down to around 1 $\mu$m  \cite{Aghion_2013}. To summarize, the use of the available spatial resolution detectors would allow to measure accelerations down to some $10^3$ g.\\
On the opposite, the probing of the fringe pattern with a third grid requires only a counting detector with a relatively low spatial distribution in order to distinguish between the $2^3 S$ atoms crossing the last grating (and annihilating on a following stopper) and the ones annihilating on it (see Fig.6). While, as seen in section 3, the first and the second gratings should be standing waves to maximize the transparency of the device, the last grid can be a physical grating with the same period because also the stopped atoms can be detected and used as confirmation of the signal generated by the crossing ones (see Ref.\cite{Brand} for the current precision in the fabrication of physical gratings and Ref.\cite{PhysRevA.96.063604,alignment} for the importance of the angular alignment of the gratings). For example, the J-PET detector, with its demonstrated 6 mm of axial spatial resolution for 2$\gamma$ annihilation events, would be able to distinguish between Ps annihilating on the $3^{rd}$ grating and on the stopper \cite{Kowalski_2018} without the need to introduce a long distance between the two elements \cite{note2}. In order to maximize the detection efficiency, a multi-layer configuration should be considered. 

\begin{figure}[htp]
\centering
\includegraphics[width=1 \linewidth]{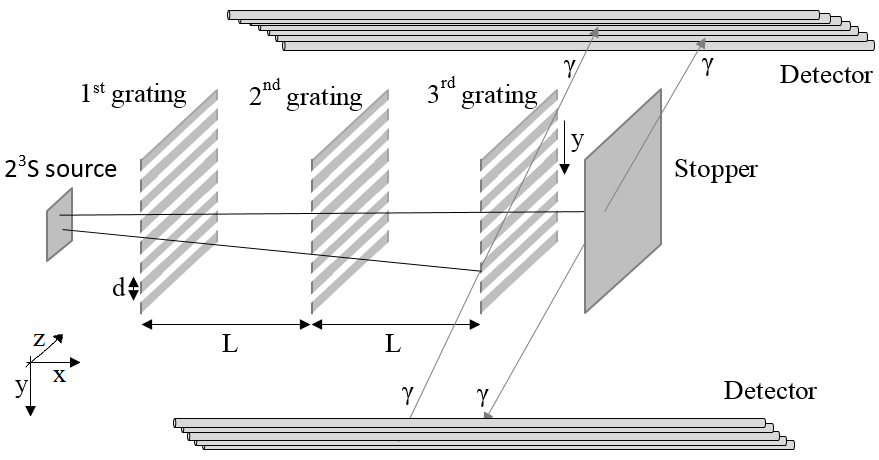}
\caption{
Possible scheme of a three gratings interferometer. Both $2^3 S$ positronium atoms passing the $3^{rd}$ grid and stopped by it emits two gamma rays that can be recorded by the integrating detector. The distribution of crossing (stopped) particles can be measured as a function of the third grid position y.}
\label{Fig.6}
\end{figure}

The conceptually easiest measurement scheme consists in moving the third grating along the force direction (y-direction in Fig.6) to scan the fringe pattern. In this configuration, sub-nm accuracy in the positioning of the third grating can be guaranteed by commercial single-axis linear piezo nanopositioning systems [see for example \cite{piezo-nanopositioners}]. This makes it feasible to resolve also $\Delta$y shifts $<$ 1 $\mu$m and measure accelerations from $10^3$ g to g. The probability of passage (and stopping) can thus be measured as a function of y (p(y)). Still following this measurement scheme, the p(y) curve has to be measured both in presence and in absence of the investigated force. In the case of the gravity the “no-force” case can be obtained by tilting the interferometer by $\pi$/2 \cite{PhysRevA.54.3165}. The main difficulties that one is expected to face following this approach are investigated in Ref.\cite{Colella}. The $\Delta$y shift given by the considered force can then be resolved by comparison of the probability distributions p(y) in the two cases. 
Given the present characteristics of quasi-monochromatic $2^3 S$ Ps beams, the scan of the entire p(y) would be very time consuming, especially for small accelerations, as each single point will require the measurement time reported in Tab.2. However, as shown in Ref.\cite{phd_braunig:14}, the scan of the pattern is not an absolute necessity and, if the shape of the expected fringe pattern (or of the passing/stopping probability distribution) is known, the measurement in a fixed position with and without acceleration (along the direction y of Fig.6) could already give a good estimate of the investigated force.

\section{Improvements of the interferometer sensitivity}
\label{sec:1}

According to the considerations above, with the present $2^3 S$ production efficiency the measurement of accelerations ranging between $10$ g and g on Ps would require, with a Mach-Zehnder interferometer and with a third grating detection scheme, between 2.6x2 weeks and 60x2 months of continuous data taking. A shortening of the measurement time is highly desirable, especially for small accelerations, and looks feasible by increasing the intensity of the $2^3 S$  beam. Further improvements to the sensitivity can come from the use of light gratings with different periodicity with respect to the 1312 nm originally proposed in Ref.\cite{oberthaler_ps:02}.


\subsection{Laser collimation of the $2^3 S$ beam}
\label{sec:2}

An important boost to the intensity of the $2^3 S$ Ps beam could come from Ps laser cooling. Although Ps laser cooling has been discussed for many years \cite{IIJIMA2000104,KUMITA2002171}, it has not yet been experimentally demonstrated even if the atomic structure of Ps does not present any fundamental limitations to Ps cooling \cite{Shu_2016,cassidy_review:18}. 
Indeed, the most efficient transition in terms of the energy removed from the system per cooling cycle is the $1^3 S-2^3 P$ transition; therefore a 1D laser cooling scheme requires a long duration counter-propagating $\sim$243 nm laser beam crossing the expanding Ps cloud in the direction in which the cooling is needed \cite{IIJIMA2000104}. In order to collimate the Ps beam, two laser beams, perpendicular to each other and perpendicular to the axial diffusion direction of Ps, have to be employed. The duration of these pulses should be limited to few tens of ns. Indeed, as demonstrated in Ref.\cite{PhysRevA.99.033405}, a longer delay between the positron implantation in the $e^+$/Ps converter and the excitation to $3^3 P$ level implies a reduction of the $2^3 S$ production efficiency. However, Monte Carlo simulations reported in Ref.\cite{pauline}, show that this time appears sufficient to cool an important fraction of the Ps in the laser direction. 
There, 25 ns long laser pulses with 5 kW power, 50 GHz spectral bandwidth and 75 GHz and 120 GHz detunings from the $1^3 S-2^3 P$ transition resonance where considered. When applied to the 1D Doppler distribution of $\sim$1000 K Ps (like in \cite{aegis_neq3:16}), the effect of 75 GHz laser pulses is to increase by a factor 1.75 the amount of Ps travelling with a velocity  lower than $\sim 10^4$ m/s or to enhance by a factor 1.2 the amount of Ps travelling with a velocity  lower than $\sim 2.5 * 10^4$ m/s if the 120 GHz detuned pulse is used. 
As the average modulus of the velocity of Ps at 1000 K is of the order of $10^5$ m/s, the atoms with a velocity component along the laser beam below $\sim 10^4$ m/s and $\sim 2.5 * 10^4$ m/s expand with an angle lower than $\pm7^{\circ}$ and $\pm14^{\circ}$ with respect to the axial direction of the beam, respectively. If the two laser beams are employed simultaneously with the 75 GHz-detuned laser beam perpendicular to the long side of the slit (vertical direction in Fig.1) and the 120 GHz-detuned laser perpendicular to the first one (horizontal direction in Fig.1), the number of atoms expanding in the solid angle defined by the slit  increases by $1.75*1.2\sim$2.1 times. Thus, the following excitation to $2^3 S$ level will result in a 2.1 times more intense beam reaching the interferometer (see Tab.3).

\begin{table*}
\centering
\caption{Expected enhacements in the beam intensity and in the sensitivity of the interferometer given by laser collimation of the beam, use of slower Ps, of a Raman excitation scheme and use of light gratings with periodicity of 730/2 nm. The effects on the measurement time to perform tests of g and 10 g are reported.} 
\label{tab:3}
\begin{tabular}{cccccc}
\toprule
Beam/interferometer    &              angular  & $2^3 S$ at the entrance                             & Needed shots  to reach   & notes  \\
interferometer              &          acceptance& of the device                                               &                                          &\\
characteristics              &                             &  per shot                                                      &$a_{min} = g   \qquad      (10 g)$&\\
\midrule
\midrule
Present $2^3S$ beam  & 5.5 mrad & $\sim$680	                & $\sim$4 $10^6$    \qquad     ($\sim$4 $10^4)$           & 	                    \\
 &&&$\sim$60 months   \quad    (2.6 weeks)& \\
\midrule
Laser collimation	           &5.5 mrad	  & $\sim$1430	                & $\sim$1.9 $10^6$ \qquad ($\sim$1.9 $10^4)$      	& x2.1 beam intensity \\
&&           &$\sim$29 months \quad (1.2 weeks)&  with 2D laser collimation\\
\midrule
Use of slower  $2^3S$  &	11 mrad&$\sim$2860	&$\sim$ $10^6$ \qquad ($10^4$)&     The considered doubling of the angular    \\
 along  &&& $\sim$14.3 months \quad (4.3 days)&acceptance can be obtained by halving\\
the beam direction &&&& the Ps velocity (from $10^5$ to 5 $10^4$ m/s)\\

\midrule
Raman scheme	&11 mrad	&$\sim$10780	& $\sim$2.6 $10^5$      \qquad   ($\sim$2.6 $10^3)$  	&x3.77 beam intensity \\
&&&$\sim$3.8 months    \quad    (1.2 days)&\\
\midrule
 n=2$\rightarrow$n=17 	&11 mrad & $\sim$10780 & $\sim$ 8 $10^4    \qquad     (\sim$ 8 $10^2 )$  & x1.8 in the sensitivity \\
d=365 nm &&&$\sim$5 weeks  \quad    (8.5 hours)& of the interferometer\\
 &&&& corresponding to a reduction of $1.8^2$\\
 &&&&in the measurement time\\
\midrule
 n=2$\rightarrow$n=17 	&20 mrad & $\sim$19400 & $\sim$ 4.5 $10^4      \qquad    (\sim$ 4.5 $10^2)$  & x1.8 in the angular \\
d=365 nm &&&$\sim$2.8 weeks   \quad     (4.7 hours)& acceptance and consequently\\
&&&& in the beam intensity\\
\midrule
\bottomrule
\end{tabular}
\end{table*}

\subsection{Use of slower $2^3 S$ along the beam direction}
\label{sec:2}

The dependency of both the deflection induced by a given acceleration (Eq.(2)) and the minimum detectable acceleration (Eq.(3) and Eq.(4)) from the interrogation time $t$ -determined by the lifetime of the test particle- looks to close any possibility to increase the beam intensity by acting on the $2^3 S$ speed. Indeed, every change in the $2^3 S$ velocity along the beam direction requires a rescale of the device length $L$ to keep $t$ in the optimum detection range (see Fig.4). However, the reduction of the velocity has the advantage of increasing the de Broglie wavelength of the test particle (Eq.(1)) and, as a consequence, to enhance the diffraction angle in the Mach-Zehnder interferometer (Eq.(6)). This allows working with a larger angular acceptance of the beam. For instance, the halving of the Ps speed from $10^5$ m/s to 5 $10^4$ m/s, keeping the same intensity of the $2^3 S$ source and halving the dimensions of the device (the distance of the first grating from the source and the length $L$), doubles the angular acceptance of the beam and thus its intensity (see Tab.3).\\
The realization of quasi- monochromatic $2^3 S$ Ps with velocity lower than $10^5$ m/s has been already achieved by increasing the delay of the $1^3 S \rightarrow 3^3 P$ laser pulse; i.e. addressing Ps that has spent more time in the nanochannels \cite{PhysRevA.99.033405}. Unfortunately, with this procedure, the amount of the produced $2^3 S$ decreases by increasing the pulse delay. This problem could be circumvented by sending the  $1^3 S \rightarrow 3^3 P$ laser pulse without further delays (20 ns from the positron implantation in the target) but keeping the  $e^+$/Ps converter at a temperature lower than room temperature. Indeed, the production of slower $2^3 S$ Ps without sacrificing the produced amount has been observed in similar $e^+$/Ps converters held at cryogenic temperatures \cite{mariazzi2010prl,TOF_MC,velocimetry}.

\subsection{Raman excitation scheme}
\label{sec:2}

Stimulated decay from $3^3 P$ level allows to have a $\sim 4.5 \%$ efficiency of $2^3 S$ production with respect to the overall Ps emitted into vacuum by the converter (See Sec.2). The use of a more efficient monochromatic $2^3 S$ production scheme would permit to enhance the beam intensity with important benefits for interferometry measurements. 
A promising alternative to the incoherent two-step $1^3 S \rightarrow 3^3 P \rightarrow 2^3 S$ is a coherent Raman excitation scheme involving an intermediate virtual level [see for example \cite{Foot}]. In order to achieve the required transition, the Raman excitation scheme could use simultaneously a 212 nm laser pulse ($5^{th}$ YAG harmonic) and a 1700 nm pulse. The advantage of this scheme is that in Raman excitation absorption and stimulated emission occur simultaneously. Thus, a $\pi$-pulse transfers all the addressable population from the $1^3 S$ level to $2^3 S$ \cite{Foot} without the distribution of the population on the different involved levels typical of the two-step scheme. 
The addressable fraction of Ps that one could excite via coherent Raman excitation is determined by the limited geometrical overlap of the laser spots on the Ps cloud and the large Doppler profile of Ps emitted from the target that is not entirely covered by the limited spectral width of the lasers. As shown in Ref.\cite{aegis_neq3:16}, the first step ($1^3 S \rightarrow 3^3 P$) of the two-step transition has an efficiency equal to  $\sim 16 \%$ of the overall population of Ps emitted by the target. As the maximum theoretical $1^3 S \rightarrow 3^3 P$ efficiency is expected to be $\sim 93 \%$, it means that the addressable fraction of Ps is 16/93$\sim$17$\%$ \cite{aegis_neq3:16}. Thus, if in the Raman scheme, laser pulses with similar band width and spot dimension to the lasers used in Ref.\cite{aegis_neq3:16} are employed, up to 17$\%$ of the overall Ps could be transferred to the $2^3 S$ level. This would be an increment of 3.77 times in the $2^3 S$ beam intensity with respect to the $1^3 S \rightarrow 3^3 P \rightarrow 2^3 S$ two-step procedure (Tab.3). 

\subsection{Light gratings with periodicity smaller than 1312 nm}
\label{sec:2}
 
The sensitivity estimations for a Mach-Zehnder interferometer reported in section 3.2 are for the use of light gratings with a wavelength of $\sim$1312 nm and a small detuning from the $2 \rightarrow 3$ transition, according to the original proposal of Ref.\cite{oberthaler_ps:02}. However, transitions from n=2 to levels higher than n=3 can be used to reduce the period of the gratings. Ultimately, the only upper limit is set by the level at which photoionization of the $2^3$S Ps starts to become effective (below $\lambda$=730 nm for Ps in n=2), killing a part of the beam. To avoid this undesirable effect one could sit close to the $2 \rightarrow 17$ transition whose energy gap is 1.679 eV, corresponding to $\sim$730 nm. If standing waves with this wavelength are used, their period is reduced to $d \sim$365 nm. According to Eq.(3), the decrease of $d$ proportionally reduces the minimum detectable acceleration. Going from $d \sim$ 1312/2 nm to $d \sim$ 730/2 nm, the sensitivity of the interferometer directly increases by a factor $\sim$1.8.
The reduction of the grating period has another positive effect on the interferometry measurement. Indeed, as pointed out by Eq.(6), the reduction of the grating period increases the diffraction angle in the Mach-Zehnder interferometer. Once again, this allows having a larger angular acceptance of the beam that could be increased up to 20 mrad. This 1.8 enhancement in angular acceptance of the beam implies an identical increase of its intensity (Tab.3).\\
The small grating period and the increased divergence of the beam in this measurement scheme will set stricter requirements on the precision of the alignement of the standing waves of light in order to preserve the visibility of the fringes. With the interferometer and the beam considered here, a visibility $ >$0.15 will require a longitudinal asymmetry of the gratings smaller than $\sim$4 $\mu$m and a rotational misalignement smaller than $\sim$20 $\mu$rad \cite{PhysRevA.96.063604}.

\section{Conclusion}
\label{sec:1}

In this work, we have analyzed the possibility to conduct force-sensitive inertial experiments on a matter/antimatter and purely leptonic system like metastable Ps in the $2^3 S$ state. Two possible inertial sensitive devices have been considered: a classical moiré deflectometer and a Mach-Zehnder interferometer. In spite of the need imposed by the Mach-Zehnder interferometer to work with a quite collimated beam, the very high transparency offered by its light gratings and their short period make the sensitivity of this device up to several tens of times better than that of a moiré deflectometer. Thus a Mach-Zehnder interferometer is preferable in the case of the measurement of very weak forces, like the gravitational interaction. 
Adopting already-demonstrated techniques for the production of a $2^3 S$  beam, the Mach-Zehnder interferometer offers the possibility to measure accelerations of the order of 100 g in a few hours and 10 g in a few weeks. This sensitivity level would be sufficient for the first observation of Ps atoms’ deflection by a laser dipole force. The measurement of acceleration of the order of g, with the present $2^3 S$ beam, would require very long data taking (60x2 months).\\
The shift of the fringe pattern due to an optical dipole force or to gravity on $2^3 S$ Ps is expected to be in the $\mu$/sub-nm range. The spatial resolution necessary to resolve the smallest shifts can be reached with a three gratings scheme where the third mechanical grating is placed in the plane where the fringes are expected. The fraction of crossing and stopped $2^3 S$ on the third grating can then be recorded in a fixed position of the grating with and without the investigated acceleration. The stopping/crossing probability can then be compared to extract the information about the strength of the investigated force.\\
Different strategies to increase the $2^3 S$ beam flux and to improve the sensitivity of the devices have been proposed and analyzed. Among them, the most promising are Ps beam collimation through 2D laser Doppler cooling in the transverse direction and the coherent Ps Raman excitation from the ground state to the $2^3 S$ level. In combination, the two techniques would increase the flux of the beam by almost a factor eight. Other improvements to the sensitivity can come from the use of slower $2^3 S$ Ps and the realization of standing wave gratings with a period of $\sim$365 nm corresponding to the $n=2 \rightarrow n=17$ transition. By implementing all these improvements, the data taking to measure the gravitational acceleration on Ps would decrease to few weeks, yielding a first test of the weak equivalence principle for a leptonic matter/antimatter system. With the same improvements, the measurement of an acceleration of 10 g (achievable by employing an optical dipole force) would require only a few hours.

\section{Authors contributions}
The original idea of this work was from RC. It was expanded togheter with SM. MD, GN and RSB gave important suggestions. All the authors were involved in the preparation of the manuscript.
All the authors reviewed and approved the manuscript.

\section{Acknowledgments}
We are grateful to P. Yzombard and D. Comparat for the useful feedbacks about laser cooling. This work was financially supported by the European Union's Horizon 2020 research and innovation programme under the Marie Sklodowska-Curie grant agreement No. 754496 - FELLINI.

%

\begin{thebibliography}{}
%
%
\bibitem{PhysRev.82.455}
M. Deutsch, Phys. Rev. \textbf{82}, (1951) 455-456.
\bibitem{Gidley}
C. L. Wang,  M.H. Weber, K.G. Lynn and K.P. Rodbell, App. Phys. Lett. \textbf{81}, (202) 4413-4415.
\bibitem{Liszkay_2012}
L. Liszkay, F. Guillemot, C. Corbel,J-P Boilot, T. Gacoin, E. Barthel, P. Perez, M-F Barthe, P. Desgardin, P. Crivelli, U. Gendotti and A. Rubbia, New J. of Phys.l \textbf{14}, (2012) 065009.
\bibitem{PhysRevLett.112.045501}
M. Zanatta, G. Baldi, R.S. Brusa, W. Egger, A. Fontana, E. Gilioli, S. Mariazzi, G. Monaco, L. Ravelli and F. Sacchetti, Phys. Rev. Lett. \textbf{112}, (2014) 5.
\bibitem{fee1993measurement}
M.S. Fee, S. Chu, Steven, A.P. Mills Jr, R.J.  Chichester, D.M. Zuckerman, E.D. Shaw and K. Danzmann,Phys. Rev. A \textbf{48}, (1993) 192.
\bibitem{KARSHENBOIM20051}
S.G. Karshenboim, Phys. Rep. \textbf{422}, (2005) 1 - 63.
\bibitem{cassidy_review:18}
D.B. Cassidy, Eur. Phys. J. D \textbf{72}, (2018) 53.
\bibitem{PhysRevLett.46.717}
A.P. Mills, Phys. Rev. Lett. \textbf{46}, (1981) 717-720.
\bibitem{NAGASHIMA201495}
Y. Nagashima, Phys. Rep. \textbf{545}, (2014) 95 - 123.
\bibitem{michishio2016observation}
K. Michishio, T. Kanai, S. Kuma, T. Azuma, K. Wada, I. Mochizuki, T. Hyodo, A. Yagishita and Y. Nagashima, Nature com. \textbf{7}, (2016) 11060.
\bibitem{Ps}
D.B. Cassidy and A.P. Mills, Nature \textbf{449}, (2007) 195.
\bibitem{PhysRevLett.93.263401}
C.H. Storry, A. Speck, D. Sage, N. Guise, G. Gabrielse, D. Grzonka, W. Oelert, G. Schepers, T. Sefzick, H. Pittner, M. Herrmann, J. Walz, T.W. H\"ansch, D. Comeau and E.A. Hessels,  Phys. Rev. Lett. \textbf{93}, (2004) 263401.
\bibitem{aegis_grav:12}
M. Doser et al., (AEgIS collaboration), Class. and Quan. Grav. \textbf{29}, (2012) 183009.
\bibitem{Perez_2012}
P. Perez and Y. Sacquin, Class. and Quan. Grav. \textbf{29}, (2012) 184008.
\bibitem{moire}
S. Aghion et al. (AEgIS collaboration), Nature com. \textbf{5}, (2014) 4538.
\bibitem{ALPHA}
P. Hamilton, A. Zhmoginov, F. Robicheaux, J. Fajans, J. S. Wurtele, and H. M\"uller Phys. Rev. Lett. \textbf{112}, (2014) 121102 
\bibitem{mills_leventhal:02}
A.P. Mills jr and M. Leventhal, Nucl. Instr. Meth. Phys. Res. B \textbf{192}, (2002) 102-106.
\bibitem{oberthaler_ps:02}
M. Oberthaler, Nucl. Instr. Meth. Phys. Res. B \textbf{192}, (2002) 129.
\bibitem{ANANDAN1989347}
J. Anandan, Phys. Lett. A \textbf{138}, (1989) 347 - 352.
\bibitem{dipole_force}
R. Grimm, M.  Weidemuller and Y.B. Ovchinnikov, Adv. in Atom. Mol., and Opt. Phys. \textbf{42}, (2000) 95 - 170.
\bibitem{cassidy2012efficient}
D.B. Cassidy, T.H. Hisakado,  H.W.K.  Tom and A.P. Mills Jr, Phys. Rev. Lett. \textbf{108}, (2012) 043401.
\bibitem{aegis_neq3:16}
S. Aghion et al. (AEgIS collaboration), Phys. Rev. A \textbf{94}, (2016) 012507.
\bibitem{Jones_2016}
A.C.L. Jones, T.H. Hisakado, H.J. Goldman, H.W.K. Tom and A.P. Mills, J. of Phys. B: Atom., Mol. and Opt. Phys. \textbf{49}, (2016) 064006.
\bibitem{PhysRevA.95.033408}
A.M. Alonso, S.D. Hogan and D.B. Cassidy,  Phys. Rev. A \textbf{95}, (2017) 033408.
\bibitem{aegis_meta:18}
S. Aghion et al. (AEgIS collaboration), Phys. Rev. A \textbf{98}, (2018) 013402.
\bibitem{PhysRevA.99.033405}
C. Amsler et al. (AEgIS collaboration), Phys. Rev. A \textbf{99}, (2019) 033405.
\bibitem{arXiv:1904.09004}
M. Antonello et al. (AEgIS collaboration),  Phys. Rev. A, \textbf{100}, (2019) 063414.
\bibitem{PhysRevA.46.5696}
T.J. Murphy and C.M. Surko, Phys. Rev. A \textbf{46}, (1992) 5696--5705.
\bibitem{danielson2015}
J.R. Danielson, D.H.E. Dubin, R.G. Greaves, C.M. Surko, Rev. Mod. Phys. \textbf{87}, (2015) 247.
\bibitem{cassidy2006accumulator}
D.B. Cassidy, S.H.M. Deng, R.G. Greaves, A.P. and Mills Jr, Rev. of Sci. Inst. \textbf{77}, (2006) 073106.
\bibitem{aegis_nimb:15}
S. Aghion et al. (AEgIS collaboration), Nucl. Instr. Meth. Phys. Res. B \textbf{362}, (2015) 86-92.
\bibitem{mariazzi2010prb}
S. Mariazzi, P. Bettotti, S. Larcheri, L. Toniutti, L and R.S. Brusa, Phys. Rev. B \textbf{81}, (2010) 235418.
\bibitem{mariazzi2010prl}
S. Mariazzi, P. Bettotti and R.S. Brusa, Phys. Rev. Lett. \textbf{104}, (2010) 243401.
\bibitem{PhysRevA.54.3165}
M.K. Oberthaler, S. Bernet, E.M. Rasel, J. Schmiedmayer and A. Zeilinger, Phys. Rev. A \textbf{54}, (1996) 3165-3176.
\bibitem{phd_braunig:14}
P.H.M. Br{\"a}unig,  \textit{Atom Optical Tools for Antimatter Experiments} (PhD thesis, Ruperto-Carola University of Heidelberg, Heidelberg, 2014).
\bibitem{PhysRevA.96.063604}
A. Demetrio, S.R. M\"uller, P. Lansonneur and M.K. Oberthaler, Phys. Rev. A \textbf{96}, (2017) 063604.
\bibitem{Colella}
R. Colella, A.W. Overhauser, and S.A. Werner, Phys. Rev. Lett. \textbf{34}, (1975) 1472.
\bibitem{Baudon}
J.  Baudon, R. Mathevet and J. Robert, J. Phys. B: At. Mol. Opt. Phys. \textbf{32}, (1999) R173
\bibitem{Miffre}
A. Miffre, M. Jacquey, M. B\"uchner, G. Trenec, and J. Vigue, Phys. Scr.  \textbf{74}, (2006) C15
\bibitem{Cronin}
A.D. Cronin, J. Schmiedmayer, and D. E. Pritchard, Rev. Mod. Phys. \textbf{81}, (2009) 1051
\bibitem{note}
Such a scheme could be more difficult for Rydberg states. Indeed, the light crystal needs a wavelength very close to a given transition and for Rydberg states the adjacent levels are very close. For instance, for Ps in n=15, the standing light wave for the Mach-Zehnder interferometer should be of the order of 340 $\mu$m (corresponding to the gap with n=16). If one decides to use a different transition (for instance n=2-n=15) the possibility of photoionization of the level has to be taken into account.
\bibitem{Giltner}
D.M. Giltner, R.W. McGowan, and S.A. Lee, Phys. Rev. Lett. \textbf{75}, (1995) 2638 
\bibitem{MCP}
C. Amsler et al. (AEgIS collaboration), Nucl. Instrum. Methods Phys. Res., Sect. B \textbf{457}, (2019) 44-48.
\bibitem{PACIFICO201612}
N. Pacifico et al. (AEgIS collaboration), Nucl. Instrum. Methods Phys. Res., Sect. A \textbf{831}, (2016) 12-17.
\bibitem{Tremsin_2018}
A.S. Tremsin, J.V. Vallerga, R.R Raffanti, Journal of Instrumentation \textbf{13}, (2018) C11005--C11005.
\bibitem{Aghion_2013}
S. Aghion et al. (AEgIS collaboration), J. of Inst. \textbf{8}, (2013) P08013-P08013.
\bibitem{Brand}
C. Brand, M. Sclafani, C. Knobloch, Y. Lilach, T. Juffmann, J. Kotakoski, C. Mangler, A. Winter, A. Turchanin, J. Meyer, O. Cheshnovsky, and M. Arndt, Nature nanotech. \textbf{10}, (2015) 845
\bibitem{alignment}
A. Miffre, M. Jacquey, M. B\"uchner, G. Trenec, and J. Vigue, Eur. Phys. J. D \textbf{33}, (2005) 99–112
\bibitem{Kowalski_2018}
P. Kowalski et al. (J-PET collaboration), Phys. in Med. and Bio. \textbf{63}, (2018) 165008.
\bibitem{note2}
The short distance between $3^{rd}$ grating and the stopper is required to minimize the number of self-annihilations in between. In the case of  $ 2^3 S $ atoms with a velocity of $ 10^5 m/s $, the number of self-annihilations in the indicated distance of 6 mm is less the $5 \%$. 
\bibitem{piezo-nanopositioners}
https://www.aerotech.co.uk/product-catalog/piezo-nanopositioners/qnp-l-series.aspx; https://www.pi-usa.us/en/products/piezo-flexure-nanopositioners/x-linear-piezo-flexure-nanopositioning-stages/p-752-high-precision-nanopositioning-stage-200800/.
\bibitem{IIJIMA2000104}
H. Iijima, T. Asonuma, T. Hirose, M. Irako, T. Kumita, M. Kajita, K. Matsuzawa and K. Wada, Nucl. Instrum. Methods Phys. Res., Sect. A \textbf{455}, (2000) 104-108.
\bibitem{KUMITA2002171}
T. Kumita, T. Hirose, M. Irako, K. Kadoya, B. Matsumoto, K. Wada, N.N. Mondal, H. Yabu, K. Kobayashi and M. Kajita, Nucl. Instrum. Methods Phys. Res., Sect. B \textbf{192}, (2002) 171 - 175.
\bibitem{Shu_2016}
K. Shu, X. Fan, T. Yamazaki, T. Namba, S. Asai, K. Yoshioka and M. Kuwata-Gonokami, J. of Phys. B: Atom., Mol. and Opt. Phys. \textbf{49}, (2016) 104001
\bibitem{pauline}
P. Yzombard, \textit{Laser cooling and manipulation of antimatter in the AEgIS experiment} (PhD thesis, Université Paris-Saclay, Paris, 2016).
\bibitem{TOF_MC}
F. Guatieri, S. Mariazzi, L. Penasa, R.S. Brusa, G. Nebbia and C. Hugenschmidt, submitted to Phys. Rev. X, (2019).
\bibitem{velocimetry}
M. Antonello et al. (AEgIS collaboration), submitted to Phys. Rev. A, (2019).
\bibitem{Foot}
C.J. Foot, \textit{Atomic Physics}, Chap. 9, (Oxford University Press, Oxford 2005).
\end{thebibliography}
%

\end{document}